# Interaction between magnetic moments and itinerant carriers in d0 ferromagnetic SiC

Yu Liu,[1,*] Ye Yuan,[1,6] Fang Liu,[1,6] Roman Böttger,[1] Wolfgang Anwand,[2] Yutian Wang,[1] Anna Semisalova,[1,7] Alexey N. Ponomaryov,[3] Xia Lu,[4] Alpha T. N'Diaye,[5] Elke Arenholz,[5] Viton Heera,[1] Wolfgang Skorupa,[1] Manfred Helm,[1,6] Shengqiang Zhou[1]

[1]Helmholtz-Zentrum Dresden-Rossendorf, Institute of Ion Beam Physics and Materials Research, 01328 Dresden, Germany

[2]Helmholtz-Zentrum Dresden-Rossendorf, Institute of Radiation Physics, 01328 Dresden, Germany

[3]Helmholtz-Zentrum Dresden-Rossendorf, Dresden High Magnetic Field Laboratory (HLD-EMFL), 01328 Dresden, Germany

[4]College of Energy, Beijing University of Chemical Technology, Beijing 100029, China

[5]Advanced Light Source, Lawrence Berkeley National Laboratory, Berkeley, California 94720, USA

[6]Technische Universität Dresden, 01062 Dresden, Germany

[7]Lomonosov Moscow State University, Faculty of Physics, 119991 Moscow, Russia

[*]E-mail: y.liu@hzdr.de

Elucidating the interaction between magnetic moments and itinerant carriers is an important step to spintronic applications. Here, we investigate magnetic and transport properties in d0 ferromagnetic SiC single crystals prepared by post-implantation pulsed laser annealing. Magnetic moments are contributed by the *p* states of carbon atoms but their magnetic circular dichroism is different from that in semi-insulating SiC samples. The anomalous Hall Effect and negative magnetoresistance indicate the influence of d0 spin order on free carriers. The ferromagnetism is relatively weak in N implanted SiC compared with that in Al implanted SiC after annealing. The results suggest that d0 magnetic moments and itinerant carriers can interact with each other, which will facilitate the development of SiC spintronic devices with d0 ferromagnetism.

In a common sense, local magnetic moments indispensable for ferromagnetism come from elements with partially filled 3d or 4f subshells. High-spin states are guaranteed according to the Hund's rules in these elements. Then, spontaneous spin coupling will favor ferromagnetic order if complying with Stoner criterion, that is, its exchange energy multiplying the density of states is larger than unity [1, 2]. The exchange integral is sensitive to the distance between moments but coupling across several atoms is still possible as itinerant carriers can play a role as intermedium, known as the RKKY theory [3-5]. This theory has further led to the discovery of giant magnetoresistance and has brought the great success of spintronics [6]. Similarly, this interaction can also exist in diluted magnetic semiconductors. The interaction between carriers and spins provides the possibility to fabricate devices based on spintronics [7].

Defect-induced ferromagnetism is an intriguing phenomenon as no transition metal directly donates moments in the system. A lot of efforts have been given to understand this phenomenon (for recent reviews, please refer to Refs. [8] and [9]), since this unexpected ferromagnetism has been observed in highly oriented pyrolytic graphite (HOPG) and $HfO_2$ [10, 11]. There are two kinds of materials with defect-induced ferromagnetism. One has $d$ electrons in its ground state, so defects may make spin coupling obey the Stoner criterion and then induce magnetism. Materials such as $LaAlO_3/SrTiO_3$ [12, 13], $TiO_2$ [14], ZnO [15], $MoS_2$ [16] and V-doped SiC [17] belong to that category. They may also overcome the Stoner criterion [1, 2] using charge transfer, which has been demonstrated at the interfaces between metal atom layers and C60 molecules [18]. Their anomalous Hall Effect and Kondo Effect have been revealed [19, 20], so they could become promising candidates in device applications in the future. The other kind is totally different. It comprises materials, such as graphite [10], graphene [21, 22], Si [23] and SiC [24-27], without any $d$ electrons. As $s$ and $p$ electrons are less localized than $d$ electrons, the appearance of ferromagnetism is very elusive and challenging to detect. It has been found that this kind of ferromagnetism cannot be enhanced by volume [28]. Several reports have pointed out that *p* states and hydrogen-mediated electronic states of carbon vacancies can lead to ferromagnetism [29-32]. The direct evidence of grain boundary induced ferromagnetism has been further demonstrated on the surface of HOPG [33]. Moreover, the formation of local magnetic moments by removal of carbon atoms has been illustrated on a graphite surface [34] and on graphene [35].

Revealing the role of carriers is an important step to the spintronic application of SiC with defect-induced magnetism. In this context, we investigate the interaction between moments and charges in d0 ferromagnetic SiC single crystals by introducing both free carriers and defects via Al or N ion implantation and post pulsed laser annealing. The shift in magnetic circular dichroism, the anomalous Hall Effect and the different magnetic behaviors between p-type and n-type samples suggest that d0 magnetic moments and itinerant carriers can interact with each other in SiC.

## MATERIALS AND METHODS

**Materials.** A commercially available 4 inch semi-insulating 4H-SiC (000-1) wafer from Cree was cut into pieces with dimensions of $10 \times 10 \times 0.33$ mm$^3$ for performing ion implantation. All the main magnetic impurities are ruled out as they cannot be detected by secondary ion mass spectrometry (typically $< 10^{15}$ cm$^{-3}$ or even less). After implantation, they were further cut into pieces of $5 \times 5 \times 0.33$ mm$^3$ for pulsed laser annealing.

**Implantation and Annealing.** Implantation was performed at room temperature using a Danfysik 200 kV ion implanter in the ion beam center at Helmholtz-Zentrum Dresden-Rossendorf (HZDR). The incident energies for Al and N were 180 keV and 110 keV, respectively, and the incidence angle was 7º (to avoid ion channeling). Samples were implanted with fluences from $5 \times 10^{13}$ to $5 \times 10^{16}$ cm$^{-2}$ for Al and $1 \times 10^{14}$ to $1 \times 10^{17}$ cm$^{-2}$ for N, respectively. The Al and N ion implantation induced defect profiles, predicted using TRIM (transport of ions in matter) simulation, are demonstrated in Fig. S1 [36]. Each piece was irradiated only once and pristine pieces taken from the same wafer were kept as reference. Pulsed laser annealing was performed in ambient air with a 308 nm excimer laser with pulse duration of about 30 nanoseconds. The samples for annealing have the fluences of $1 \times 10^{16}$ cm$^{-2}$ for Al and $5 \times 10^{16}$ cm$^{-2}$ for N, respectively. The energy density for the implanted SiC is 0.2, 0.4, 0.6, 0.8 J/cm$^{-2}$, respectively.

**Characterization.** Raman spectroscopy was carried out on a LabRAM (HORIBA Jobin Yvon) using a 532 nm Nd:YAG laser as the excitation source with a resolution of ~ 0.5 cm$^{-1}$. The positron annihilation Doppler broadening spectroscopy (PADBS) was applied to clarify the nature of defects created by ion implantation and modified due to laser annealing. PADBS measurements were carried out with the mono-energetic slow positron beam ''SPONSOR'' at HZDR [37] with the positron energy E from 30 eV to 36 keV. The energy resolution of the Ge detector at 511 keV is (1.09 ± 0.01) keV, resulting in a high sensitivity to changes in material properties from surface to depth of several µm. The S parameter (shape, central/total count ratio) is mainly a measure for the size and the concentration of the open volume in the material, while the W parameter (wing, tail/total count ratio) describes the chemical surrounding area of the open volume. Electron spin resonance (ESR) was performed using Bruker ELEXSYS-II EPR spectrometer operating at X-band microwave frequency (9.4 GHz) at room temperature. The derivative of absorption was recorded during magnetic field sweeping. The magnetization properties were measured by using the vibrating sample magnetometer attached with a commercial superconducting quantum interference device (SQUID-VSM, Quantum Design) with a sensitivity limit of $10^{-7}$ emu. During the ferromagnetic resonance (FMR) measurement, the forward amplitude of complex transmission coefficients (S21) was output with a coplanar wave guide and recorded by a vector network analyzer (Agilent E5071C). The temperature and field control was realized by means of a Physical Properties Measurement System (Quantum Design) [38]. The samples were attached to the coplanar wave guide with insulating silicon paste to guarantee good contact. X-ray magnetic circular dichroism (XMCD) spectroscopy measurements were performed at BL 6.3.1 of the Advanced Light Source using total electron yield (TEY) detection in magnetic fields of 0.5 T at 79 K for Si K-edge and 300 K for C K-edge, respectively. The field was reversed for each photon energy while the X-ray polarization remained constant. Transport properties were measured using a current of 1 mA under magnetic fields (within 50 kOe) and temperatures controlled by a Lakeshore system.

**RESULTS AND DISCUSSION**

Al or N impurities are the best for p or n-type doping in SiC [39, 40]. Therefore, we choose to implant them into SiC to introduce free holes/electrons, which can also realize defect-induced ferromagnetism at the same time. The results of the Al doped samples are demonstrated first and those of N doped will be discussed later. We first examined the crystallinity of the samples after implantation. Figure S2(a) shows that all Raman modes in the intact lattice are gone after Al implantation with the fluence of $1 \times 10^{16}$ cm$^{-2}$ [36], so much damage was introduced into the SiC substrates [40]. However, no ferromagnetism is found in the as-implanted samples. For example, the magnetization as a function of magnetic field in the samples with Al fluences of $1 \times 10^{15}$ cm$^{-2}$, $1 \times 10^{16}$ cm$^{-2}$, $5 \times 10^{16}$ cm$^{-2}$ shows no hysteresis in Fig. S2(b) [36].

After pulsed laser annealing, Raman peaks of the folded transverse and longitudinal acoustic (FTA and FLA), the folded transverse and longitudinal optic (FTO and FLO) modes emerge again, showing that part of damage has been removed [Fig. 1(a)]. In the 0.2 J/cm$^{-2}$ sample, the FTA mode is missing as the energy is not high enough to recover the lattice. The intensity of its Raman peaks is relatively weak and the background noise is stronger than that of the rest samples. The spectra of annealed 4H-SiC crystals with the laser energy densities of 0.4 and 0.6 J/cm$^{-2}$ are similar to that of the pristine one [41], and no other SiC polytypes or secondary phases in these samples are detected within the sensitivity of Raman characterization. A Raman peak of Si appears in the sample annealed by the energy density of 0.8 J/cm$^{-2}$, suggesting that this energy density is too high to keep stability of SiC during annealing.

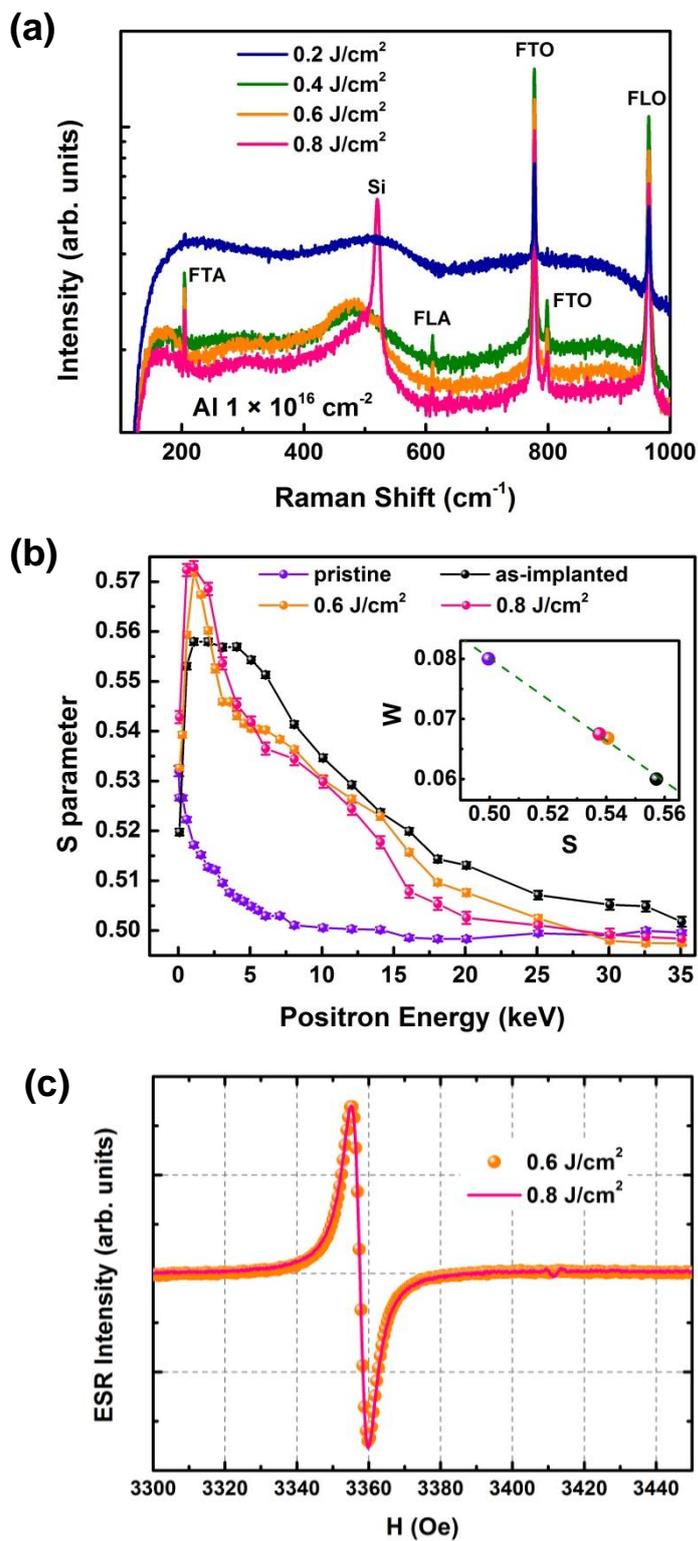

FIG. 1. The structural information of SiC samples. (a) The Raman spectra for Al implanted SiC samples with the fluence of $1 \times 10^{16}$ cm$^{-2}$ after pulsed laser annealing with the energy density from 0.2 to 0.8 J/cm$^2$. (b) The S parameter vs. incident positron energy for pristine, Al implanted without annealing, and Al implanted followed by annealing samples in the PADBS. The fluence of Al implantation is $1 \times 10^{16}$ cm$^{-2}$. The energy density of pulsed laser annealing is 0.6 and 0.8 J/cm$^2$. Inset: The corresponding S-W plot. (c) The ESR spectra of the 0.6 and 0.8 J/cm$^2$ samples at 300 K.

According to our previous positron annihilation experiments [25, 26, 28, 42, 43], divacancies have been identified in neutron or ion irradiated SiC. They also dominate defects after annealing [43]. Figure 1(b) shows the measured S parameters vs. the incident positron energies. There are two annihilation states detected in the pristine sample: a surface state at low positron energies and a bulk state $S_{bulk} = 0.5$ from the positron energy E > 8 keV, representing the S parameter for 4H-SiC without defects. The S parameter of the as-implanted sample inside the implantation region (between 1 and 4 keV) is larger than that of the pristine SiC which points to vacancy-type defects created by ion implantation. The defects in the implantation region beneath the surface of the as-implanted sample have a size of about 10 divacancies based on estimation according to a scaling curve for defects in 6H-SiC [44]. This vacancy clustering could be responsible for the absence of ferromagnetism after implantation [26]. The S parameter of samples after annealing considerably increases at the surface as the defects are driven from the implantation region to the surface and form bigger clusters after the high temperature treatment. Consequently, the size of defects in the rest implanted region of these samples is reduced (S changes from 0.56 to 0.54), which becomes similar to that in ferromagnetic Ne implanted SiC in the previous report [26]. Moreover, the linear S-W relation in the inset of Fig. 1(b) confirms that defects in all SiC samples are the same, i.e., agglomerations of Si-C divacancies [44, 45]. In ESR measurements in Fig. 1(c), it is further confirmed the dominating role of divacancies in defects after annealing, as the resonance is located at around 3359 Oe corresponding to the g factor 2 with the narrow linewidth of 5 Oe [46].

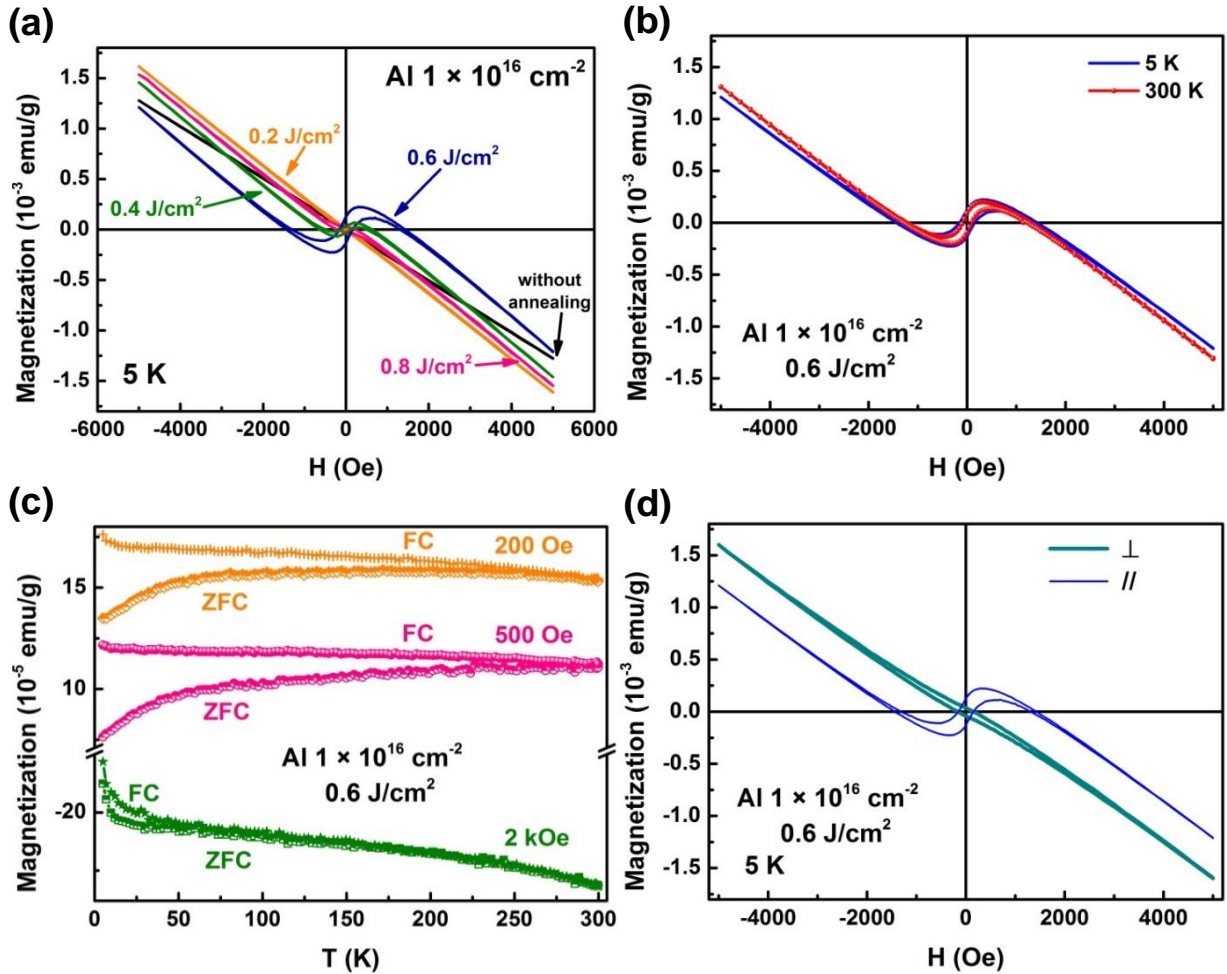

FIG. 2. The magnetic properties of Al implanted SiC after pulsed laser annealing. (a) The magnetization as a function of magnetic field within 5 kOe at 5 K. (b) The magnetization vs. field for the sample annealed with the energy density of 0.6 J/cm$^2$ at 5 K and 300 K for comparison. (c) The zero field cooling (ZFC) / field cooling (FC) magnetization as a function of temperature from 5 K to 300 K under magnetic fields of 200, 500 and 2000 Oe. (d) The magnetization as a function of magnetic field measured with the field both perpendicular and parallel to the sample surface (000-1) within 5 kOe. Clear magnetic anisotropy is observed.

Figure 2(a) shows the magnetization vs. magnetic field in the range of 5 kOe at 5 K. Before annealing, only diamagnetic features can be observed in the as-implanted samples. After annealing with a low energy 0.2 J/cm$^2$, a weak hysteresis loop can be observed in a small field range. With increasing pulsed laser energies, the magnetic order gradually enhances. A distinct hysteresis loop with ferromagnetic features can be clearly seen and the saturation magnetization reaches $5 \times 10^{-4}$ emu/g in the sample annealed using 0.6 J/cm$^2$. Then, the saturation in the 0.8 J/cm$^2$ decreases compared with that in 0.6 J/cm$^2$, which shows the rise and fall feature of defect-induced magnetism reported in previous works on SiC [26] and Si [23]. At 300 K, the diamagnetism increases slightly while the hysteresis loop is similar as shown in Fig. 2(b). So the room-temperature ferromagnetism is induced by post-implantation annealing. The absence of tiny parasitic or any secondary magnetic phase is confirmed by the smooth and featureless magnetization vs. temperature curve [Fig. 2(c)]. Weak paramagnetism due to irradiation damage is revealed when the field of 2 kOe is applied. Figure 2(d) shows that the hysteresis measured with fields both perpendicular and parallel to the sample surface (000-1), i.e. C face, at 5 K. The in-plane direction is found to be the easy axis. In the isosurface spin density plot of a previous report [25], the ab-plane spin coupling is shown stronger than along c axis, which could be the reason of this magnetic anisotropy. However, considering that the implanted layers are thin, the phenomenon may also be attributed to the shape anisotropy. Overall, the anisotropy corroborates that the magnetism is intrinsic, as foreign contamination or clusters are unlikely to bring either structural or shape anisotropy.

To understand the magnetism in annealed SiC and investigate its origin, we carried out the FMR and XMCD measurements. Figure S3 demonstrates the FMR frequency vs. magnetic field mapping for the 0.6 J/cm$^2$ laser annealed SiC at 300 K [36]. The shift of the resonance valley is not linear, so the signal is not only from the contribution of free electrons. The valley broadens when the external field is reduced, which is brought by exchange interaction between moments [38]. It provides convincing evidence on the ferromagnetism in Al implanted SiC after pulsed laser annealing. The XMCD spectroscopy can probe ferromagnetism with element-specificity by using circularly polarized X rays and tuning the photon energy to absorption edges characteristic for certain elements, i.e. ~1850 eV and 285 eV for K-edges of Si and C, respectively, exciting electrons from the 1$s$ level to unoccupied and possibly spin polarized $p$ states [47, 48]. Figure 3 shows typical X-ray absorption spectra (XAS) and XMCD signals obtained from an Al implanted SiC sample after annealing. The data at the silicon K-edge was obtained at 79 K and at the carbon K-edge at 300 K, respectively. There is no detectable XMCD signal at the silicon K-edge, while an XMCD is clearly seen at the carbon K-edge. Though SiC samples are doped by Al in this case, the basic mechanism behind the magnetism of these samples should be the same as that of neutron irradiated / noble gas ion implanted SiC because the created defects are all dominated by divacancies. Therefore, the magnetic moments mainly come from the $p_z$ orbitals at nearest-neighbor carbon atoms of defect sites according to previous reports [31, 32]. It is also noticed that two peaks at ~286 and ~290 eV appear in the XMCD spectrum at the carbon K-edge. The corresponding peak of Xe implanted (undoped) samples is located at ~284 eV [32] and its position is ~291 eV in Al doped SiC polycrystalline [31]. So both cases exist in Al implanted samples after annealing. The emergence of the peak at ~290 eV shows that the distribution of local moments changes after holes are introduced. The shift from 286 eV to 290 eV means that $p^*$ states in higher energy levels can also form local moments, so introducing p-type carriers are able to increase the number of $p$ states contributing to ferromagnetism in SiC, which suggests that the p-type doping can be expected to enhance defect-induced magnetism.

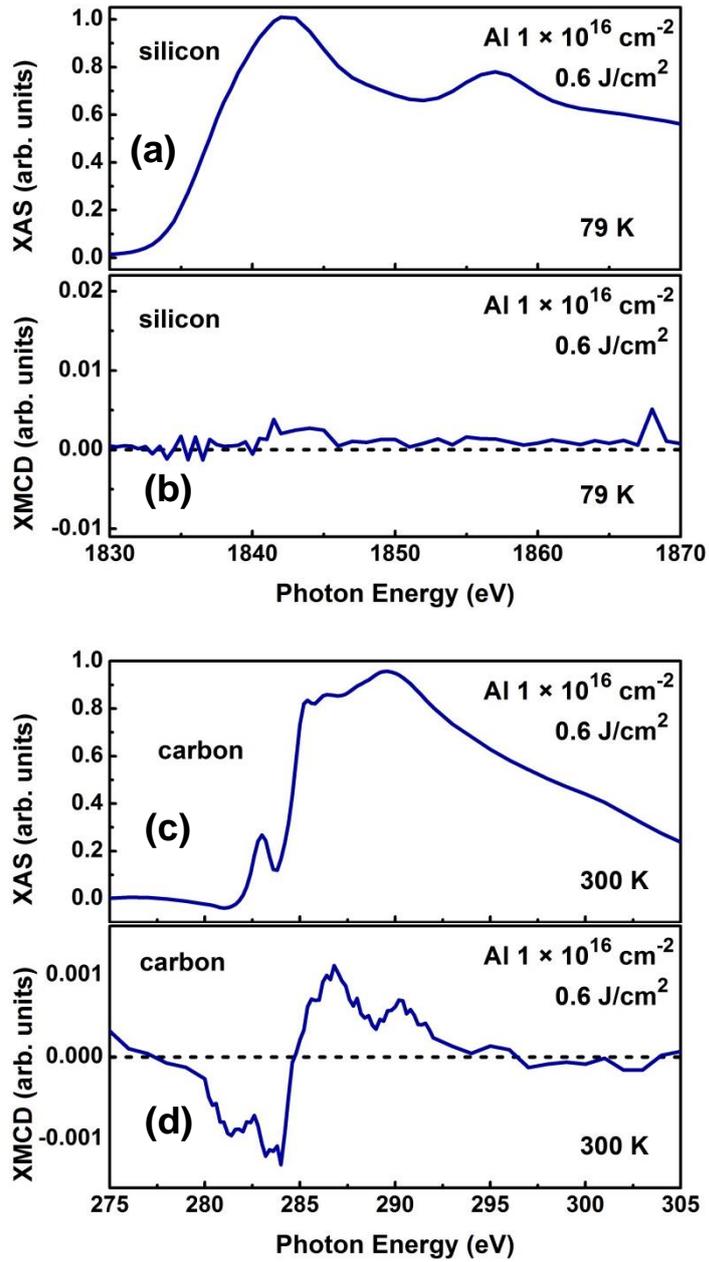

FIG. 3. XAS and XMCD of Al implanted SiC after 0.6 J/cm² annealing at the silicon K-edge at 79 K and at the carbon K-edge at 300 K. (a) XAS at the Si K-edge, (b) XMCD at the Si K-edge, (c) XAS at the C K-edge, and (d) XMCD at the C K-edge.

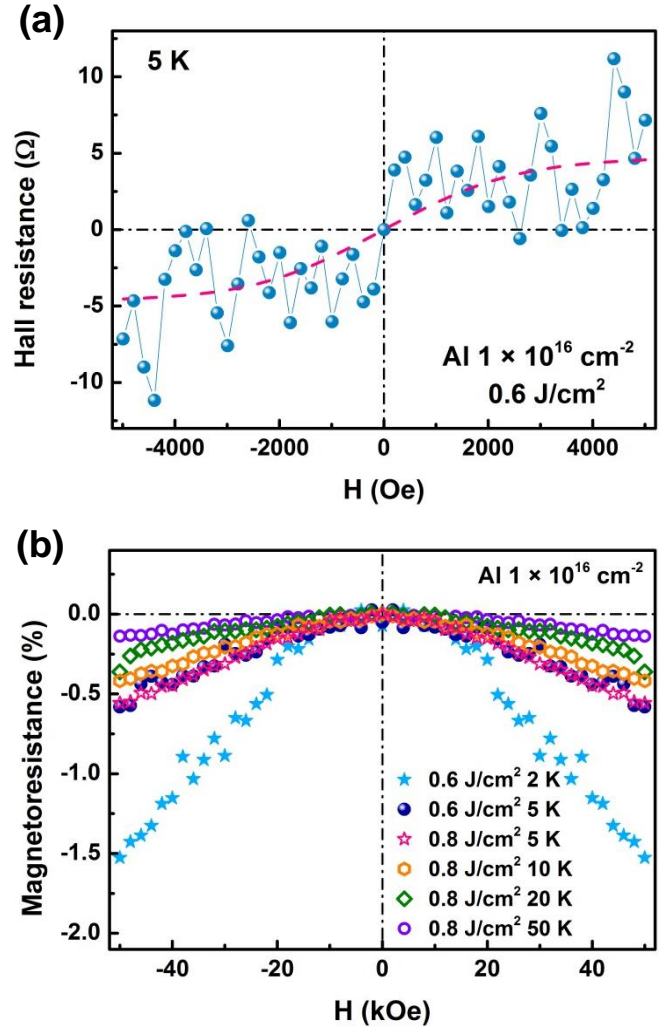

FIG. 4. The transport properties of Al implanted SiC after pulsed laser annealing. (a) The Hall resistivity of the 0.6 J/cm$^2$ sample with the field of 5 kOe at 5 K. The pink dashed line is a guide to the eye. (b) The magnetoresistance of the 0.6 J/cm$^2$ and 0.8 J/cm$^2$ in the field range of 50 kOe.

In the Hall effect measurement, the concentration and the mobility of carriers in SiC implanted with the Al fluence of $1 \times 16$ cm$^{-2}$ after 0.6 J/cm$^2$ pulsed laser annealing are estimated to be $3.4 \times 10^{19}$ cm$^{-3}$ and 0.40 cm$^2$/(Vs), respectively, while they are $2.3 \times 10^{20}$ cm$^{-3}$ and 0.02 cm$^2$/(Vs) for the 0.8 J/cm$^2$ sample. In the low field range of 5 kOe, the trace of anomalous Hall Effect is revealed in Fig. 4(a). The weak and noisy signal is probably due to the low carrier mobility. Weak negative magnetoresistance is observed in the samples annealed with 0.6 J/cm$^2$ and 0.8 J/cm$^2$ shown in Fig. 4(b). The magnetoresistance of the 0.6

J/cm$^2$ sample can reach -1.5% under 50 kOe at 2 K, while at 5 K it is almost the same as that of the 0.8 J/cm$^2$ sample. As the magnetization of the 0.6 J/cm$^2$ sample is larger than that of 0.8 J/cm$^2$ sample, it implies that impurities and defects also contribute to the negative magnetoresistance. The negative magnetoresistance weakens when the temperature increases, and vanishes above 50 K.

In contrast to Al implanted SiC, ferromagnetism is relatively difficult to be induced in N implanted SiC. No ferromagnetism is found as shown in Fig. S4(a) before pulsed laser annealing, similar to that of Al implanted SiC [36]. After pulsed laser annealing, only the 0.2 J/cm$^2$ sample shows ferromagnetism with the saturation magnetization of $1.6 \times 10^{-4}$ emu/g in Fig. 5. The magnetism almost remains the same at 300 K [see Fig. S4(b)] [36]. Dopants need proper energy density (around 0.6 J/cm$^2$ in this case) to be activated, so 0.2 J/cm$^2$ is not sufficient to activate the most of N dopants. Higher energy density means higher carrier concentrations. Therefore, the absence of ferromagnetism in the samples with higher energy annealing suggests that electron doping can suppress the ferromagnetism. Although the previous reports point out the possibility for negative charged defects to induce ferromagnetism [25, 28], our calculations confirm that further increasing negative charge can eliminate spin polarization, which is consistent with the experimental results.

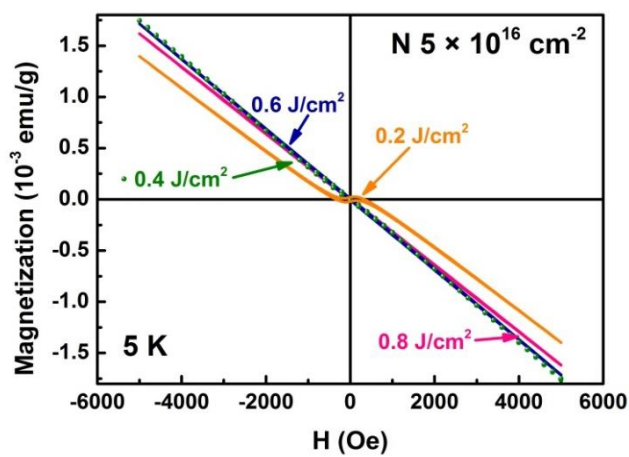

FIG. 5. The magnetization as a function of magnetic field within 5 kOe for N $5 \times 10^{16}$ cm$^{-2}$ implanted SiC samples after pulsed laser annealing at 5 K.

Based on the investigation above, it is confirmed that carriers play a crucial role in defect-induced magnetism. First, the XMCD peak shifts due to the different carrier concentrations, so the origin of magnetism can be affected by carrier concentrations. More *p* states could have the possibility to contribute local moments after introducing p-type carriers. Second, the anomalous Hall Effect and negative magnetoresistance show that the internal magnetic field from spin ordering can influence the behavior of carriers. So the interaction exists between moments induced by divacancies and carriers due to doping. However, the anomalous Hall Effect is almost lost in the noise. It is understood as that the defect-induced magnetism is weak and the electron mobility in SiC is low. Last but not least, in contrast to p-type SiC, ferromagnetism is relatively difficult to be induced in n-type SiC. This phenomenon is common in diluted magnetic semiconductors, which is elucidated that the electrons in p-type semiconductors are more localized than those in n-type semiconductors [7, 49, 50]. This similarity implies that, the theory developed for diluted magnetic semiconductors could be borrowed to describe defect induced magnetic materials.

So far we have investigated the role of carriers in Al/N ion implanted SiC after pulsed laser annealing. The other techniques like carrier injection, gating or optical pumping could be better options to accurately control the carrier concentrations. It is expected to further approach the application of defect-induced magnetism in SiC or other semiconductors, such as spin transistors [51], spin solar cells [52], spin light emitting diodes [53], and spin amplifiers [54].

**CONCLUSIONS**

In conclusion, we have investigated magnetic and transport properties in SiC single crystals implanted with Al or N ions followed by pulsed laser annealing treatment. The major defects are confirmed to be divacancies. No ferromagnetism is found in as-implanted samples even though much damage is introduced, while intrinsic ferromagnetism emerges after the pulsed laser annealing changes the distribution of defects. The rise and fall of saturation magnetization as a feature of defect-induced ferromagnetism is revealed in the samples after annealing, and the Al implanted sample annealed at 0.6 J/cm$^2$ can have the saturation magnetization of $5 \times 10^{-4}$ emu/g. The magnetic anisotropy is observed with the in-plane direction as the easy axis. The magnetic moments are mostly contributed from the $p_z$ orbitals at nearest-neighbor carbon atoms of defect sites but their magnetic circular dichroism is different from that of semi-insulating SiC samples. The anomalous Hall Effect and the negative magnetoresistance reveal the influence of defect-induced spin order on carriers. Different from that in p-type samples, the ferromagnetism in N implanted SiC is only observed after 0.2 J/cm$^2$ pulsed laser annealing. High electron concentrations are found to suppress the defect-induced magnetism in SiC. The results confirm the interaction between magnetic moments and itinerant carriers in SiC with defect-induced ferromagnetism. These findings will facilitate the design and optimization of SiC spintronic devices utilizing d0 ferromagnetism.

**ACKNOWLEDGEMENTS**

Y.L. would like to thank Dr. Yuelei Zhao from Peking University for the FMR measurements, Ning Liu, Dr. Di Gan, Dr. Jiao Huang, Ying Tian and Prof. Yang Sun from Institute of Physics, CAS for the assistance in the experiments. The work is financially supported by the Helmholtz Postdoc Programme (Initiative and Networking Fund, PD-146). Support by the Ion Beam Center (IBC) at HZDR is gratefully acknowledged. The Advanced Light Source is supported by the Director, Office of Science, Office of Basic Energy Sciences, of the U.S. Department of Energy under Contract No. DE-AC02-05CH11231. X.L. thanks the funding support for State Key Laboratory of Organic-Inorganic Composites (oic-201701011).

**REFERENCES**

[1]     E. C. Stoner, P. Roy. Soc. Lond. A Mat. **165**, 372 (1938).

[2]     E. C. Stoner, P. Roy. Soc. Lond. A Mat. **169**, 339 (1939).

[3]     M. A. Ruderman and C. Kittel, Phys. Rev. **96**, 99 (1954).

[4]     T. Kasuya, Prog. Theor. Phys. **16**, 45 (1956).

[5]     K. Yosida, Phys. Rev. **106**, 893 (1957).

[6]     M. N. Baibich, J. M. Broto, A. Fert, F. N. Van Dau, F. Petroff, P. Etienne, G. Creuzet, A. Friederich, and J. Chazelas, Phys. Rev. Lett. **61**, 2472 (1988).

[7]     T. Dietl, H. Ohno, F. Matsukura, J. Cibert, and D. Ferrand, Science **287**, 1019 (2000).

[8]     P. Esquinazi, W. Hergert, D. Spemann, A. Setzer, and A. Ernst, IEEE Trans. Magn. **49**, 4668 (2013).

[9]     S. Zhou, Nucl. Instrum. Meth. B **326**, 55 (2014).

[10]    P. Esquinazi, D. Spemann, R. Höhne, A. Setzer, K. H. Han, and T. Butz, Phys. Rev. Lett. **91**, 227201 (2003).

[11]    M. Venkatesan, C. B. Fitzgerald, and J. M. D. Coey, Nature **430**, 630 (2004).

[12]    A. Brinkman *et al.*, Nat. Mater. **6**, 493 (2007).

[13]    Ariando *et al.*, Nat. Commun. **2**, 188 (2011).

[14]    S. Zhou *et al.*, Phys. Rev. B **79**, 113201 (2009).

[15]    J. B. Yi *et al.*, Phys. Rev. Lett. **104**, 137201 (2010).

[16]    S. Mathew *et al.*, Appl. Phys. Lett. **101**, 102103 (2012).

[17]    H. Wang, C.-F. Yan, H.-K. Kong, J.-J. Chen, J. Xin, and E.-W. Shi, Appl. Phys. Lett. **101**, 142404 (2012).

[18]   F. A. Ma'Mari *et al.*, Nature **524**, 69 (2015).

[19]   Y. Lee, C. Clement, J. Hellerstedt, J. Kinney, L. Kinnischtzke, X. Leng, S. D. Snyder, and A. M. Goldman, Phys. Rev. Lett. **106**, 136809 (2011).

[20]   M. Lee, J. R. Williams, S. Zhang, C. D. Frisbie, and D. Goldhaber-Gordon, Phys. Rev. Lett. **107**, 256601 (2011).

[21]   R. R. Nair *et al.*, Nat. Commun. **4**, 2010 (2013).

[22]   L. L. Chen, L. W. Guo, Z. L. Li, H. Zhang, J. J. Lin, J. Huang, S. F. Jin, and X. L. Chen, Sci. Rep. **3**, 2599 (2013).

[23]   Y. Liu *et al.*, Phys. Rev. B **94**, 195204 (2016).

[24]   B. Song *et al.*, J. Am. Chem. Soc. **131**, 1376 (2009).

[25]   Y. Liu, G. Wang, S. C. Wang, J. H. Yang, L. A. Chen, X. B. Qin, B. Song, B. Y. Wang, and X. L. Chen, Phys. Rev. Lett. **106**, 087205 (2011).

[26]   L. Li, S. Prucnal, S. D. Yao, K. Potzger, W. Anwand, A. Wagner, and S. Zhou, Appl. Phys. Lett. **98**, 222508 (2011).

[27]   X. He, J. Tan, B. Zhang, M. Zhao, H. Xia, X. Liu, Z. He, X. Yang, and X. Zhou, Appl. Phys. Lett. **103**, 262409 (2013).

[28]   Y. Wang *et al.*, Phys. Rev. B **92**, 174409 (2015).

[29]   H. Ohldag, T. Tyliszczak, R. Hohne, D. Spemann, P. Esquinazi, M. Ungureanu, and T. Butz, Phys. Rev. Lett. **98**, 187204 (2007).

[30]   H. Ohldag, P. Esquinazi, E. Arenholz, D. Spemann, M. Rothermel, A. Setzer, and T. Butz, New J. Phys. **12**, 123012 (2010).

[31]   M. He, X. He, L. Lin, B. Song, and Z. H. Zhang, Solid State Commun. **197**, 44 (2014).

[32]   Y. T. Wang *et al.*, Sci. Rep. **5**, 8999 (2015).

[33]   J. Cervenka, M. I. Katsnelson, and C. F. J. Flipse, Nat. Phys. **5**, 840 (2009).

[34]   M. M. Ugeda, I. Brihuega, F. Guinea, and J. M. Gómez-Rodríguez, Phys. Rev. Lett. **104**, 096804 (2010).

[35] Y. Zhang *et al.*, Phys. Rev. Lett. **117**, 166801 (2016).

[36] See Supplemental Material at [URL will be inserted by publisher].

[37] W. Anwand, G. Brauer, M. Butterling, H. R. Kissener, and A. Wagner, in *Defect and Diffusion Forum* (Trans Tech Publ, 2012), pp. 25.

[38] Y. Zhao, Q. Song, S.-H. Yang, T. Su, W. Yuan, S. S. Parkin, J. Shi, and W. Han, Sci. Rep. **6**, 22890 (2016).

[39] M. V. Rao, J. Tucker, O. W. Holland, N. Papanicolaou, P. H. Chi, J. W. Kretchmer, and M. Ghezzo, J. Electron. Mater. **28**, 334 (1999).

[40] V. Heera, D. Panknin, and W. Skorupa, Appl. Surf. Sci. **184**, 307 (2001).

[41] S. Nakashima and H. Harima, Phys. Stat. Sol. A **162**, 39 (1997).

[42] G. Brauer, W. Anwand, P. G. Coleman, A. P. Knights, F. Plazaola, Y. Pacaud, W. Skorupa, J. Störmer, and P. Willutzki, Phys. Rev. B **54**, 3084 (1996).

[43] G. Brauer, W. Anwand, P. G. Coleman, J. Störmer, F. Plazaola, J. M. Campillo, Y. Pacaud, and W. Skorupa, J. Phys.: Condens. Matter **10**, 1147 (1998).

[44] W. Anwand, G. Brauer, and W. Skorupa, Appl. Surf. Sci. **194**, 131 (2002).

[45] G. Brauer *et al.*, Phys. Rev. B **74**, 045208 (2006).

[46] N. T. Son *et al.*, Phys. Rev. Lett. **96**, 055501 (2006).

[47] G. Schütz, W. Wagner, W. Wilhelm, P. Kienle, R. Zeller, R. Frahm, and G. Materlik, Phys. Rev. Lett. **58**, 737 (1987).

[48] J. Stöhr, H. A. Padmore, S. Anders, T. Stammler, and M. R. Scheinfein, Surf. Rev. Lett. **05**, 1297 (1998).

[49] T. Dietl, H. Ohno, and F. Matsukura, Phys. Rev. B **63**, 195205 (2001).

[50] T. Dietl and H. Ohno, Rev. Mod. Phys. **86**, 187 (2014).

[51] C. Betthausen, T. Dollinger, H. Saarikoski, V. Kolkovsky, G. Karczewski, T. Wojtowicz, K. Richter, and D. Weiss, Science **337**, 324 (2012).

[52]  B. Endres, M. Ciorga, M. Schmid, M. Utz, D. Bougeard, D. Weiss, G. Bayreuther, and C. H. Back, Nat. Commun. **4**, 2068 (2013).

[53]  G. Kioseoglou and A. Petrou, J. Low Temp. Phys. **169**, 324 (2012).

[54]  Y. Puttisong, I. A. Buyanova, A. J. Ptak, C. W. Tu, L. Geelhaar, H. Riechert, and W. M. Chen, Adv. Mater. **25**, 738 (2013).